\documentclass[a4paper]{jpconf}
\usepackage[latin2]{inputenc}
\usepackage{t1enc}
\usepackage{graphicx}
\usepackage{cite}
\usepackage{hyperref}
\usepackage{amssymb,amsmath,epsfig,bm,pifont}
\newcommand{\be}{\begin{equation}}\newcommand{\ee}{\end{equation}}%
\newcommand{\bd}{\begin{displaymath}}\newcommand{\ed}{\end{displaymath}}
\newcommand{\bit}{\begin{itemize}}                                        
 \newcommand{\eit}{\end{itemize}}                                         
\newcommand{\ben}{\begin{enumerate}}                                      
 \newcommand{\een}{\end{enumerate}}                                       
\newcommand{\baa}{\begin{array}{lll}}                                     
 \newcommand{\eaa}{\end{array}}                                           
\newcommand{\ba}{\begin{eqnarray}}                                        
 \newcommand{\ea}{\end{eqnarray}}                                         
\newcommand{\Ds}{\displaystyle}                                           
\newcommand{\gev}[1]{\relax\ifmmode{\text{GeV}^{#1}}                      
                     \else{GeV$^{#1}${ }}\fi}                             
\def\MSbar{\relax\ifmmode\overline                                        
            {\rm MS}\else{$\overline{\rm MS}${ }}\fi}                     
\def\as{\relax\ifmmode \alpha_s\else{$ \alpha_s${ }}\fi}                  
\def\abar{\relax\ifmmode{\bar{a}}\else{$\bar{a}${ }}\fi}                  
    
\usepackage{color}
\definecolor{mBlue}{rgb}{0,0,1}
 
\definecolor{mRed}{rgb}{1,0,0}
 
\definecolor{mGreen}{rgb}{0.0,1.0,1.0}
 
\begin{document}
\title{Nucleon structure functions in the truncated moments approach\footnote{
Presented at XVII Workshop on High Energy Spin Physics DSPIN-17, September 11-15
2017, Dubna, Russia}}

\author{D Str\'ozik-Kotlorz$^{1,\, 2}$, S V Mikhailov$^2$ O V Teryaev$^2$
and A Kotlorz$^3$}

\address{$1$ Opole University of Technology, Department of Physics,
Pr\'oszkowska 76, 45-758 Opole, Poland}

\address{$2$ Bogoliubov Laboratory of Theoretical Physics, JINR,
         141980 Dubna, Russia}

\address{$3$ Opole University of Technology, Department of Mathematics and IT
Applications, Pr\'oszkowska 76, 45-758 Opole, Poland}

\ead{\href{mailto:dorota@theor.jinr.ru}{dorota@theor.jinr.ru}}

\begin{abstract}
We demonstrate advantages of the truncated Mellin moments (TMM) approach
in the analysis of DIS data. We present a novel method for determination
of the Bjorken sum rule (BSR) from restricted in $x$ variable experimental data.
We show how to incorporate different uncertainties for each kinematic bin.
We apply our analysis to recent COMPASS data.
\end{abstract}

\section{Introduction}\label{sec:sec1}
Our understanding of the matter structure and fundamental particle interactions
at high energies is mostly provided by deep inelastic scattering (DIS)
of leptons on hadrons and hadron-hadron collisions.
Due to the property of the collinear factorization for these hard processes hadron
characteristics
can be explained in terms of the parton distribution functions (PDFs) $f_p(x,\mu^2)$
-- universal process-independent normalized distributions describing how the hadron
momentum $P$ is partitioned in $x\cdot P$ between partons of type $p$.
The hard momentum transfer for DIS is $q$, $-q^2=Q^2 \gg P^2=m^2_h$, the Bjorken variable
$x$ satisfies the condition $0< x = Q^2/(2Pq)<1$.
Non-perturbative strong interaction forms these distributions $f_p(x,\mu^2)$ at hadronic scale $m^2_h$,
but the dependence on the normalization scale $\mu^2$ is ruled in
perturbative QCD by the well-known
Dokshitzer-Gribov-Lipatov-Altarelli-Parisi (DGLAP) evolution equations
\cite{ref1}.
Otherwise, one can study how to evolve with this scale $\mu^2$ the Mellin
moments of the parton distributions $\tilde{f}_p(n,\mu^2)$ --
the integrals of PDF $\tilde{f}(n,\mu^2)= \int^1_0 f(x,\mu^2) x^{n-1} dx$.
These moments provide a natural framework of QCD analysis as they
originate from the basic formalism of operator product expansion and
are essential in testing DIS sum rules.
However, such moments, in principle cannot be extracted from any
experiment due to kinematic constraints,
 $x \geqslant x_\text{min}=Q^2_\text{min}/(2(Pq)_\text{max}) > 0$,
 inevitably appearing in real DIS with constraints $Q^2 > Q^2_\text{min}$
 and transferred energy $(Pq) < (Pq)_\text{max}$.
By this reason, it is demanded to introduce  visible observables   with a goal
to overcome these kinematic constraints.
Generalized moments of the  $f(x,\mu^2)$,
$\Ds f(z;n,\mu^2)= \int^1_z f(x,\mu^2) x^{n-1} dx$, 
include the experimentally unavoidable lower limit of integration
 $z \geqslant x_\text{min}$, and in this
way the kinematic constraint can be taken into account.
``Truncated'' Mellin moments of the parton densities in QCD
analysis was introduced in the late 1990s \cite{ref2}.
The diagonal integro-differential DGLAP-type evolution equations for
the single and double truncated moments of the parton densities were
derived in \cite{ref3} and \cite{ref4,ref5}.
The main finding was that the $n$th truncated moment of the parton density also
obeys the DGLAP equation
but with a rescaled evolution kernel $P'(z)=z^n P(z)$ \cite{ref3}.
The TMM approach has been successfully applied in spin physics to
generalize Wandzura-Wilczek relation in terms of the
truncated moments and to derive the evolution equation for the structure
function $g_2$ \cite{ref5}.
The evolution equations for cut moments and their applications to
the quark-hadron duality were discussed in \cite{ref6}.
The collinear factorization formula for DIS structure functions in terms of
TMM, together with its advantages was given in \cite{ref7}.
A valuable generalization of the TMM approach incorporating
multiple integrations as well as multiple differentiations of the
original parton distribution was developed in \cite{ref8}.
Recently, based on the TMM, we constructed a device which allows one to
improve experimental determination of the Bjorken sum rule (BSR) \cite{ref9}.
Here, we present the main pragmatic results of our recent approach \cite{ref10}
- the generalized BSR that allows effective determination
of the BSR value from the experimental data in a restricted
kinematic range of $x$.
We also show how to incorporate different uncertainties for each kinematic bin.
We apply our analysis to recent COMPASS data on the spin structure function $g_1$.

\section{Generalized Bjorken sum rule}\label{sec:sec2}
The generalized truncated moment $f(z,n,\omega)$, obtained
as a Mellin convolution of the PDF $f$ with any normalized
function $\omega(x)$,
\begin{subequations}
\label{eq.2.1}
\begin{eqnarray}
\label{eq.2.1a}
f(x,n;\omega)= \left(\omega \ast x^nf\right)&\equiv& \int_{x}^{1}\omega \left(x/z\right)
~f(z,\mu^2)\,z^{n}\,\frac{dz}{z}, \\
&&\int_{0}^{1}\omega(t) dt =1\, ,
\end{eqnarray}
 \end{subequations}
obeys the DGLAP evolution equation with the rescaled kernel
\cite{ref9}:
\begin{equation}\label{eq.2.2}
{\cal P}(y) = P(y)\cdot y^{n}.
\end{equation}
In the case of the non-singlet polarized structure function
$g_1$ and for $n=0$, one obtains the generalized structure function
$g_{1;\,\omega}$,
\begin{equation}\label{eq.2.3}
g_{1;\,\omega}(x) \equiv \left(\omega \ast g_1\right)(x)
\equiv\int_{x}^{1}\omega \left(x/z\right)
~g_1(z,Q^2)\,\frac{dz}{z}
\end{equation}
with the same DGLAP evolution kernel as $g_1$, \cite{ref10}, namely $P(y)$.
In this way, we define the truncated BSR, $\Gamma_{1}(x_0)$,
and simultaneously, the generalized cut Bjorken sum rules (gBSR),
$\Gamma_{1;\,\omega}(x_0)$,
\begin{eqnarray}
\Gamma_{1}(x_0) &=& \int_{x_0}^{1}g_1(x)\,dx\,,\label{eq.2.4a}\\
\Gamma_{1;\,\omega}(x_0) &=& \int_{x_0}^{1}g_{1;\,\omega}(x)\,dx, \label{eq.2.4b}
\end{eqnarray}
which are equal to the ordinary BSR as $x_0 \to 0$:
\begin{equation}\label{eq.2.5}
\Gamma_{1;\,\omega}(0)=\int_{0}^{1}g_{1;\,\omega}(x)\,dx
 = \int_{0}^{1} g_1(x)\, dx\equiv\Gamma_1(0).
\end{equation}
Now one can estimate the value of $\Gamma_1(0)$ from the smooth
extrapolation of the truncated moments $\Gamma_{1;\,\omega}(x_0)$ in $x_0$.
To this aim, one constructs \textit{a bunch} of different
$\Gamma_{1;\,\omega}(x_0)$ based on the simple sign-changing normalized function
$\omega (x)$ depending on three parameters $z_1,~z_2,~A$,
\begin{equation}\label{eq.2.6}
\omega (z) = -A\,\delta(z-z_1)+(1+ A)\,\delta(z-z_2).
\end{equation}
Here the $\omega$-model parameters are $z_2 >z_1> x_0> 0$ and $A>0$ for the sign change.
This leads to a ``shuffle'' of the initial function $g_1$ in $x$ variable and
the generalized truncated BSR $\Gamma_{1;\,\omega}(x_0)$
\begin{equation}\label{eq.2.7}
\Gamma_{1;\,\omega}(x_0) = \int_{x_0/z_2}^{1}\,g_1(x)\,dx +
A\,\int_{x_0/z_2}^{x_0/z_1}\,g_1(x)\,dx\, 
\end{equation}
approaches the limit $\Gamma_1(0)$ more quickly than the ordinary
$\Gamma_{1}(x_0)$ in Eq.~(\ref{eq.2.4a}).
In this way, the total BSR limit $\Gamma_1(0)$ can be determined very
effectively with the use of a few first orders of Taylor expansion:
\begin{equation}\label{eq.2.8}
\Gamma_1(0) = \Gamma_{1;\,\omega}(x_0 - x_0)=\Gamma_{1;\,\omega}(x_0) - x_0\,\Gamma'_{1;\,\omega}(x_0)
+ x_0^2\,\frac{1}{2}\,\Gamma''_{1;\,\omega}(x_0) +\cdots\, .
\end{equation}
Figs.~1 and 2 illustrate the features of $\Gamma_{1;\,\omega}$, where we plot the bunch
$\Gamma_{1;\,\omega}(x_0)$, Eq.~(\ref{eq.2.7}), for different values of $A$, including
``constant behavior'' value $A=A_{00}$, ``quasi-linear behavior'' value $A=A_{02}(\bar{x})$
and the standard truncated BSR $\Gamma_{1}(x_0)$,
Eq.~(\ref{eq.2.4a}) (thick black curve). We use the popular form of parametrization
of $g_1$ at $Q_0^2=1\,\rm{GeV^2}$,
\begin{equation}\label{eq.2.9}
g_1(x,Q_0^2) = N\cdot x^a(1-x)^b(1+\gamma x),
\end{equation}
where $a=0$ in Fig.~1 and $a=-0.4$ in Fig.~2, respectively, $b=3$,
$\gamma = 5$ and the coefficient $N$ is the norm.
In our tests, in order to obtain a smooth approach of the bunch in
the experimentally available $x$ region, we fixed $z_1=0.7$ and $z_2=0.9$.
\begin{figure}[ht]
\centering{
\begin{minipage}{0.45\textwidth}
\includegraphics[width=\textwidth]{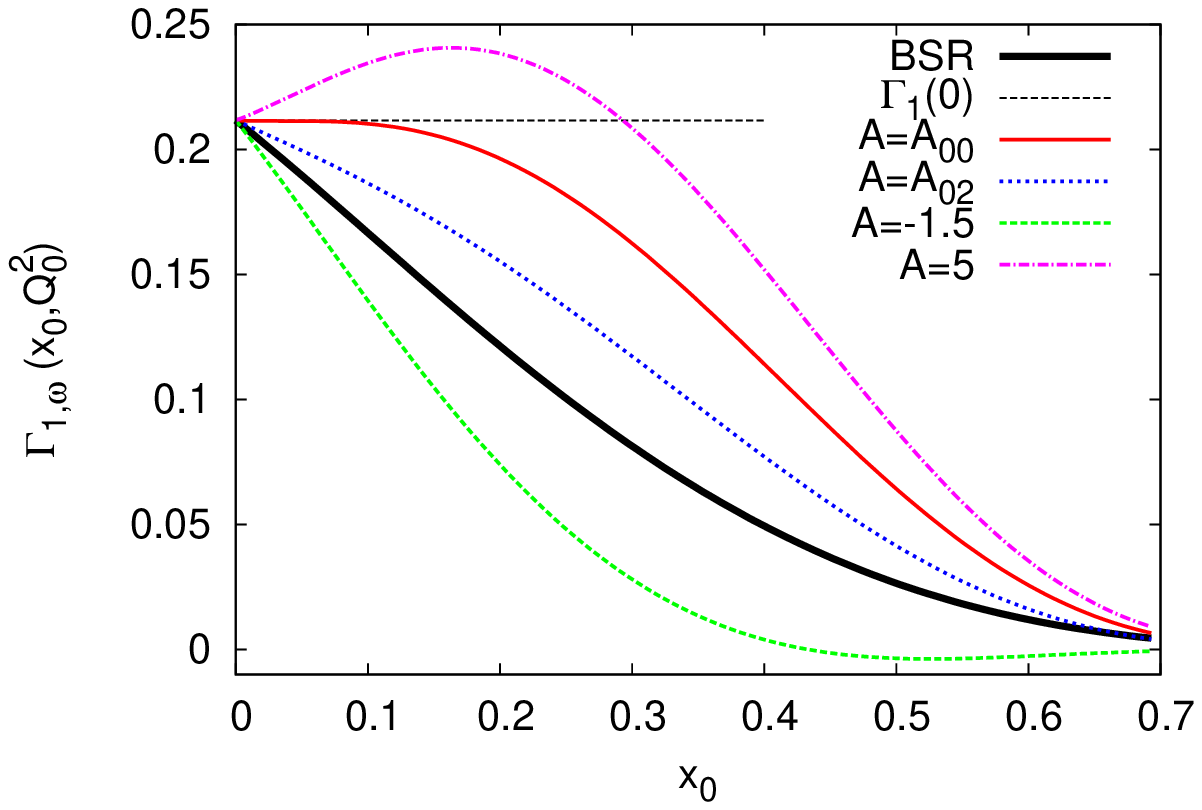}
\caption{\footnotesize \label{fig1}
$\Gamma_{1;\,\omega}(x_0)$, Eq.~(\ref{eq.2.7}), for different values of $A$
and the truncated BSR $\Gamma_{1}(x_0)$, Eq.~(\ref{eq.2.4a}) (thick black
curve) as a function of $x_0$. Input $g_1(x,Q_0^2)$, Eq.~(\ref{eq.2.9}), with $a=0$.}
\end{minipage}~~
\begin{minipage}{0.45\textwidth}
\includegraphics[width=\textwidth]{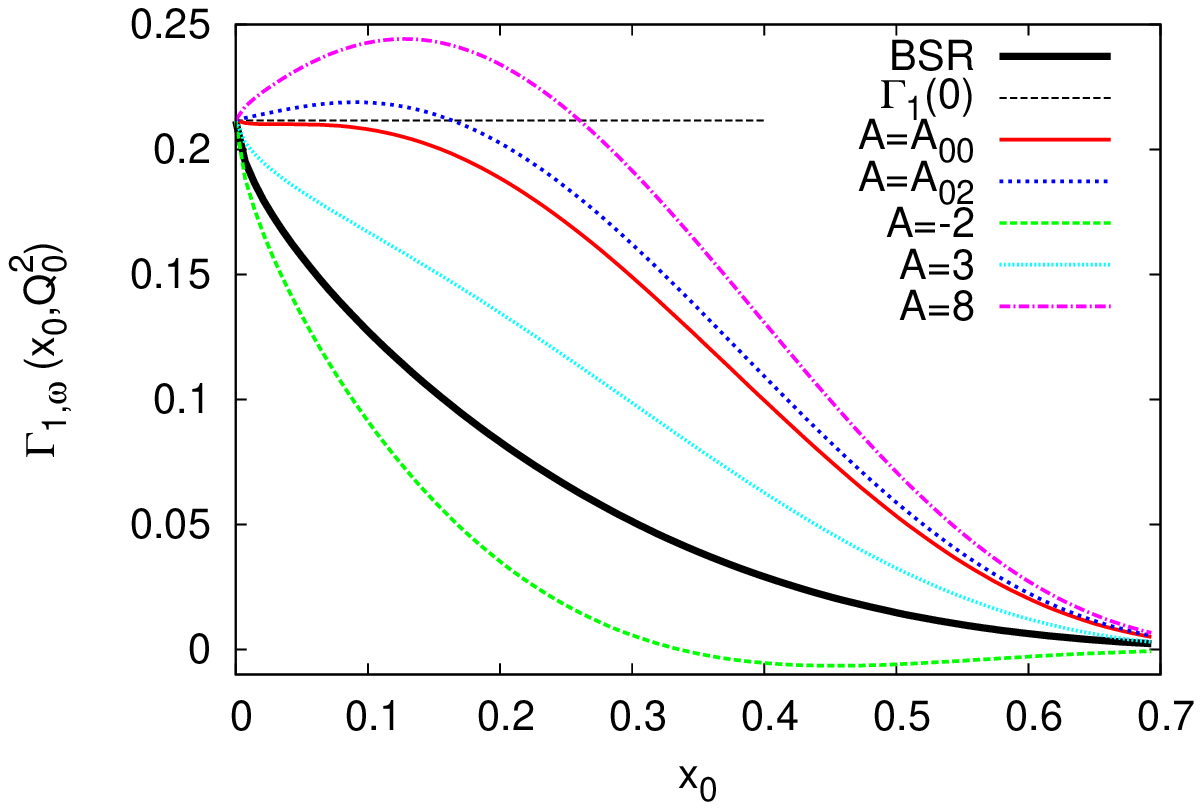}
\caption{\footnotesize \label{fig2}
$\Gamma_{1;\,\omega}(x_0)$, Eq.~(\ref{eq.2.7}), for different values of $A$
and the truncated BSR $\Gamma_{1}(x_0)$, Eq.~(\ref{eq.2.4a}) (thick black
curve) as a function of $x_0$. Input $g_1(x,Q_0^2)$, Eq.~(\ref{eq.2.9}), with $a=-0.4$.}
\end{minipage}}
\end{figure}
One can see significantly rapid saturation of $\Gamma_{1;\,\omega}(x_0)$ to
$\Gamma_{1}(0)$ in comparison with $\Gamma_{1}(x_0)$.
The quasi-linear regime of gBSR near $0$ visibly starts at rather large
values of $x_0 \gtrsim 0.1$ for the different parametrization
in Eq.~(\ref{eq.2.9}). This should ensure the applicability of the first
order approximation (IAPX),
\begin{equation}\label{eq.2.10}
\Gamma_1(0) \approx \Gamma^{\rm{IAPX}}_1(x_0) =
\Gamma_{1;\,\omega}(x_0) - x_0\,\Gamma'_{1;\,\omega}(x_0),
\end{equation}
even for JLab experimental conditions, where the admissible $x$ bunches
are rather far from 0.
In practice, one can use fit to the data instead of the ready input
parametrization. It is worthy to notice that the analysis based on the
bunch behavior allows one to shift the available region of $x$ to smaller
values, $x_0=x\cdot z_2$. In this manner, using data from large $x$ and choosing
suitable values of $z_1$ and $z_2$, one is able to get an answer in a much smaller
$x$ region.\\

The presented above idea of the generalized BSR can be used to analyze
the experimental data incorporating different uncertainties for each measurement.
Usually, the results for $g_1$ are extracted from the data for kinematic bins
$\{Q_i^2, x_i\}$ with uncertainties $\Delta(g(x_i))$.
In order to compute moments of $g_1$ and verify the BSR, bins
must be evolved to a common scale $Q^2$.
Then, the experimental bins $\{ x_i\}_1^n$ can be used to construct multi-point weight
function $\omega$ in Eq.~(\ref{eq.2.1}),
\begin{equation}\label{eq.2.11}
\omega(z) = -\sum\limits_{i=1}^{n-1}A_i\,\delta(z-z_i)+\left(1+ A\right)\,\delta(z-z_n)\,,
\end{equation}
where
\begin{equation}\label{eq.2.12}
A = \sum\limits_{i=1}^{n-1}A_i >0\,.
\end{equation}
This multi-point ($\{z_i,~A_i\}_1^{n}$) ansatz is a generalization of the
previous two-point ($\{z_1,~z_2,~A\}$) one, Eq.~(\ref{eq.2.6}), and
leads to a new presentation for $\Gamma_{1;\,\omega}$,
\begin{eqnarray}
\Gamma_{1;\,\omega}(x_0)& =&\underbrace{ \int_{x_0/z_n}^{1}\,g_1(x)\,dx +
A\int_{x_0/z_{n}}^{x_0/z_{n-1}}\,g_1(x)\,dx }_{\text{two-point result}~\Gamma_1}
+ A\,\sum\limits_{i=1}^{n-2}w_i\int_{x_0/z_{i+1}}^{x_0/z_i}\,g_1(x)\,dx.
\label{eq.2.13}
\end{eqnarray}
In order to use the generalized BSR to the determination of
$\Gamma_1(0)$, one only needs to construct the set of $\{A_i\}$.
Choosing the most appropriate weights $w_i$ and hence $A_i$,
\begin{equation}\label{eq.2.14}
w_i = \frac{1}{A}\sum\limits_{k=1}^{i}A_k\, ;~~~~~~~A_i = A\cdot(w_i-w_{i-1})\, ,
\end{equation}
one is able to tune the analysis to the real experimental constraints.
The most natural way of implementing the experimental uncertainties of $g_1$
into the ansatz of $\Gamma_{1;\,\omega}$ is to use $w_i$ inversely proportional
to the relative uncertainties $\Delta(g_1)/g_1$ at each $x_i$:
\begin{equation}\label{eq.2.15}
w_i \sim\frac{g_1(x_i)}{\Delta(g_1(x_i))}\, ,
\end{equation}
to increase the weights of the data with smaller uncertainties.

\section{Analysis of data}\label{sec:sec3}
In this section, we present practical estimation of $\Gamma_1(0)$ from
the recent COMPASS data \cite{ref11} using two-point and multi-point
versions of the gBSR approach described in the previous section.
We apply the simplified equations of our approach, written in terms
of experimental parameters.

\subsection{Two-point estimation of $\Gamma_1(0)$}\label{sec:sec3a}
For practical purposes, we write here the essential formulas for the
generalized BSR in terms of experimental data and
demonstrate the effective method for the estimation of $\Gamma_1(0)$.
Thus, the gBSR, Eq.~(\ref{eq.2.7}), where the lower limit of integration
has to be strictly related to the minimal $x$ accessible experimentally,
$x_{min}$, takes the form
\begin{equation}\label{eq.3a.1}
\Gamma_{1;\,\omega}(x_{min},r) = \int_{x_{min}}^{1}\,g_1(x)\,dx +
A\,\int_{x_{min}}^{x_{min}/r}\,g_1(x)\,dx.
\end{equation}
The experimental lower value $x_{min}$ in the above equation is related to $x_0$
from Eq.~(\ref{eq.2.7}) via
$x_0=x_{min}\cdot z_2$. The ratio parameter $r\equiv z_1/z_2$ and $x_0<x_{min}<r<1$.
We have tested all methods of estimation of $\Gamma_1(0)$ described in
\cite{ref10} and found that a very effective method, universal for
the different small-$x$ behavior of $g_1$ and for $x_{min}\lesssim 0.1$, is the
first order approximation, Eq.~(\ref{eq.2.10}).
With the use of the parameters $x_{min}$ and $r$ it reads
\begin{equation}\label{eq.3a.2}
\Gamma_1(0) \approx \Gamma^{\rm{IAPX}}_1(x_{min},r) =
\Gamma_{1;\,\omega}(x_{min},r) + (A+1)\,x_{min}\, g_1(x_{min})
- A\frac{x_{min}}{r}g_1(x_{min}/r)
\end{equation}
with $A$ requiring the second derivative of $\Gamma_{1;\,\omega}(x_0)$ to vanish
(first order approximation),
\begin{equation}\label{eq.3a.3}
A = \left[ r^2\,\frac{g'_1(x_{min}/r)}{g'_1(x_{min})}-1\right]^{-1}.
\end{equation}
Below we present our results on determination
of the BSR based on the recent the COMPASS data, where
$x_{min}=0.0036$.
We follow the method described above using
Eqs.~(\ref{eq.3a.1})--(\ref{eq.3a.3}).
Our fit to the data at $Q^2=3$~\rm{GeV}$^2$ is
\mbox{$g_1(x)\sim x^{-0.36}(1-x)^{3}(1+3.9\,x)$.}
One can see from Fig.~3 that $\Gamma_{1;\,\omega}(x)$
approaches $\Gamma_1(0)$ more quickly than the original BSR $\Gamma_1(x)$.\\

\begin{figure}[h]
\centering
\includegraphics[width=0.7\textwidth]{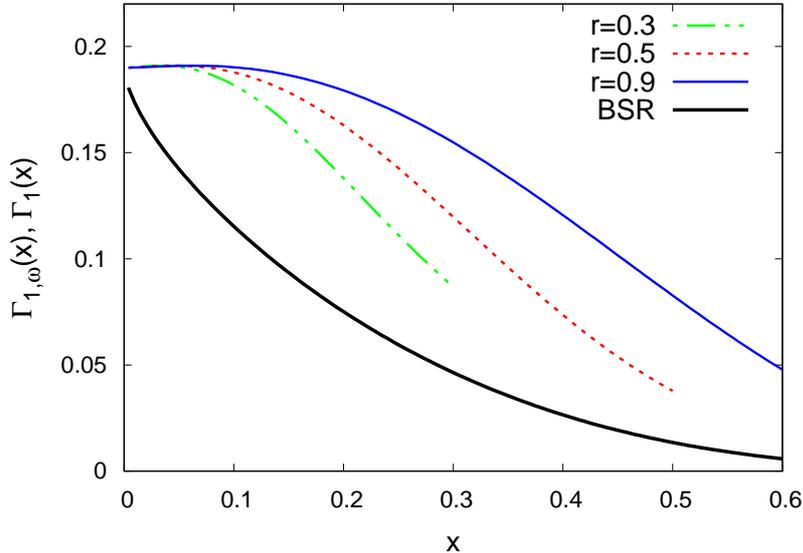}
\caption{\footnotesize
$\Gamma_{1,\omega}(x,r,Q^2)$, Eq.~(\ref{eq.3a.1}), for
$A(x_{min}=0.0036,r)$, Eq.~(\ref{eq.3a.3}) for three values of
$r: 0.9, 0.5, 0.3$, together with the truncated BSR
$\Gamma_{1}(x,Q^2)$, Eq.~(\ref{eq.2.4a}), as a function of $x$.
The results are based on our fit to the COMPASS data.
\label{fig3}}
\end{figure}
We find the estimated total BSR $\Gamma_1(0)$ from
Eqs.~(\ref{eq.3a.1})--(\ref{eq.3a.3}) for $x_{min}=0.0036$
and different $r$ (0.9, 0.5, 0.3, 0.1) $\Gamma^{\rm{IAPX}}_1(x_{min},r)=0.190$.
This should be compared to the final estimation provided by the COMPASS
collaboration,
\mbox{$\Gamma_1(0)=0.192\pm 0.007_{\rm stat}\pm 0.015_{\,\rm syst}$.}

\subsection{Multi-point estimation of $\Gamma_1(0)$}\label{sec:sec3b}
Here we develop the idea of the generalized BSR to
analyze of the COMPASS data on $g_1(x)$ incorporating
uncertainties for each measurement. We consider the results for $n$
experimental values of $x$: $x_{min}\equiv x_1<x_2<x_3<\cdots<x_n\equiv
x_{max}<1$, where $x_i=x_0/z_{n-i+1}$ and corresponding $g_1(x_i)$ together with their relative
statistical uncertainties $U_i=\Delta g_1(x_i)/g_1(x_i)$.
In terms of the experimental parameters the multi-point gBSR,
Eq.~(\ref{eq.2.13}), has the form
\begin{equation}\label{eq.3b.1}
\Gamma_{1;\,\omega}(\{ x_i\}_1^n) = \int_{x_1}^{1}\,g_1(x)\,dx +
A\,\sum\limits_{i=1}^{n-1}\tilde{w}_i\int_{x_i}^{x_{i+1}}\,g_1(x)\,dx\, ,
\end{equation}
where we choose the weights $\tilde{w}_i=U_1/U_i$ reflecting the increase of
the experimental uncertainties of $g_1$ with decreasing $x$.
The first order approximation for $\Gamma_1(0)$ in the multi-point case
is a generalization of Eqs.~(\ref{eq.3a.2}) and (\ref{eq.3a.3}):
\begin{equation}\label{eq.3b.2}
\Gamma_1(0) \approx \Gamma^{\rm{IAPX}}_1(\{ x_i\}_1^n) =
\Gamma_{1;\,\omega}(\{ x_i\}_1^n) + x_1\, g_1(x_1) +
A\,\sum\limits_{i=1}^{n-1}\tilde{w}_i \left [x_i\, g_1(x_i) - x_{i+1}\,
g_1(x_{i+1})\right ]\, ,
\end{equation}
\begin{equation}\label{eq.3b.3}
A = \frac{x^2_1\, g'_1(x_1)}{\sum\limits_{i=1}^{n-1}\tilde{w}_i \left [x^2_{i+1}\,
g'_1(x_{i+1}) - x^2_i\, g'_1(x_i)\right ]}\, .
\end{equation}
In our analysis we use the fit function to evolve $g_1(x_1,Q^2_i)$ to the
common value of $Q^2=3$~\rm{GeV}$^2$ and also to find $g'_1(x_i)$.
We obtain from Eqs.~(\ref{eq.3b.1})--(\ref{eq.3b.3})
$\Gamma_1(0) \approx \Gamma^{\rm{IAPX}}_1(\{ x_i\}_1^n)=0.190$,
the same value as in the previous case of 2-point estimation.
This result is in good agreement with the value
\mbox{$0.192\pm 0.007_{\rm stat}\pm 0.015_{\,\rm syst}$}
provided by the COMPASS \cite{ref11} from reanalyzed data.

\section{Conclusions}\label{sec:concl}
The generalized BSR based on the truncated Mellin moment
approach provides a powerful tool for testing QCD at experimental constraints.

\ack
We thank Eva-Maria Kabuss and Yann Bedfer for their help with COMPASS data.
This work is supported by the Bogoliubov-Infeld Program,
Grant No. 01-3-1113-2014/2018.
S.V.M. acknowledges support from the BelRFFR-JINR, Grant No. F16D-004.

\end{document}